\documentclass[journal]{IEEEtran}
\usepackage{amsmath,amsfonts}
\usepackage{algorithm}
\usepackage{array}
\usepackage[caption=false,font=footnotesize,labelfont=rm,textfont=rm]{subfig}
\usepackage{textcomp}
\usepackage{stfloats}
\usepackage{url}
\usepackage{verbatim}
\usepackage{graphicx}
\usepackage{cite}
\usepackage{makecell}
\usepackage{algpseudocode}
\usepackage{multirow}
\usepackage{color,xcolor}
\usepackage{microtype}
\usepackage{bbm}
\usepackage{amssymb}
\usepackage{balance}
\usepackage{amsthm}

\newcommand{\myvar}[1]{%
\usepackage[labelformat=simple]{subcaption}
\captionsetup[sub]{font=footnotesize}
\captionsetup[figure]{name={Fig.},labelsep=period,font=footnotesize} 
\renewcommand\thesubfigure{(\alph{subfigure})}
\lowercase{\def\tmp{#1}}%
\ifx\tmp x#1\else\MakeUppercase{#1}\fi}

\usepackage[colorlinks = true,
linkcolor = blue,
urlcolor  = blue,
citecolor = blue,
anchorcolor = blue]{hyperref}

\begin{document}
	
\title{Performance Analysis of Discrete-Phase Fluid Reconfigurable Intelligent Surface-Enabled Index Modulation}

\author{Peng~Zhang,~\IEEEmembership{Graduate~Student~Member,~IEEE,}
Jian~Dang,~\IEEEmembership{Senior~Member,~IEEE,}
Miaowen~Wen,~\IEEEmembership{Senior~Member,~IEEE,}
Ziyang~Liu, 
Kai-Kit~Wong,~\IEEEmembership{Fellow,~IEEE,} Chen~Zhao,~\IEEEmembership{Senior~Member,~IEEE,} Huaifeng~Shi, and Zaichen~Zhang,~\IEEEmembership{Senior~Member,~IEEE} 
\thanks{
Peng Zhang, Jian Dang, and Zaichen Zhang are with the National Mobile Communications Research Laboratory and the Frontiers Science Center for Mobile Information Communication and Security, Southeast University, Nanjing 211189, China. Jian Dang is also with the Key Laboratory of Intelligent Support Technology for Complex Environments, Ministry of Education, Nanjing University of Information Science and Technology, Nanjing 210044, China. All three authors are also with Purple Mountain Laboratories, Nanjing 211111, China (e-mail: peng\_zhang@seu.edu.cn; dangjian@seu.edu.cn; zczhang@seu.edu.cn).

Miaowen Wen is with the School of Electronic and Information Engineering, South China University of Technology, Guangzhou 510640, China (e-mail: eemwwen@scut.edu.cn).

Ziyang Liu is with the School of Communication Engineering, Hangzhou Dianzi University, Hangzhou 310018, China (e-mail: 251080010@hdu.edu.cn).

K. K. Wong is affiliated with the Department of Electronic and Electrical Engineering, University College London, Torrington Place, WC1E 7JE, United Kingdom and he is also affiliated with the Department of Electronic Engineering, Kyung Hee University, Yongin-si, Gyeonggi-do 17104, Korea.

Chen Zhao and Huaifeng Shi are with the School of Electronics and Information Engineering, Nanjing University of Information Science and Technology, Nanjing 210044, China (e-mail: 002912@nuist.edu.cn; shihuaifeng@nuist.edu.cn).

Corresponding authors: Jian Dang (e-mail: dangjian@seu.edu.cn).}
}

\markboth{IEEE Transactions on Wireless Communications,~Vol.~0, No.~0, Month~2026}%
{Zhang \MakeLowercase{\textit{et al.}}: Performance Analysis of Discrete-Phase FRIS-Enabled Index Modulation}

\maketitle
	
\begin{abstract}
  Fluid reconfigurable intelligent surfaces (FRIS) combine the position reconfigurability of fluid antennas (FA) with the phase control suburbs of reconfigurable intelligent surfaces (RIS). In this paper, we propose fluid-port-selected receive index modulation (FPS-RIM), which uses fluid-port selection as an additional spatial degree of freedom. Two operating modes are considered: FPS-RIM index-only (FPS-RIM-IO), which conveys information through the receive-antenna index, and FPS-RIM symbol-augmented (FPS-RIM-SA), which additionally conveys a constellation symbol. Under finite-resolution phase control, the active port and its quantized phase are jointly determined by the post-quantization in-phase component. A low-complexity Top-$L$ detector retains the receive-index candidates with the largest received energies and performs joint detection over the reduced candidate set. For spatially correlated FRIS cascaded channels, we derive selection-aware joint moments that account for phase quantization, fluid-port selection, and overlapping port selections between receive-index states. Correlated Gaussian distance models and quadratic-form Laplace transforms are then used to obtain unconditional pairwise error probabilities and a union-bound bit error rate expression. Numerical results show that fluid-port selection improves the error performance over conventional RIS schemes and that the proposed detector approaches maximum-likelihood performance with a moderate list size. The analytical results closely follow the simulations.
  \end{abstract}
		
	
\section{Introduction}
\label{Introduction}

\IEEEPARstart{N}{ext-generation} wireless networks are expected to support high-rate and reliable transmission for applications such as artificial-intelligence-enabled information exchange and sensor-data delivery for intelligent driving \cite{New2025FAS_Tutorial,Wu2025FAS_6G}. These requirements must be met under limited spectrum, power, and hardware resources. Spatial diversity can improve link reliability without requiring additional spectrum, but conventional fixed-position antennas cannot adapt their locations to spatial fading.

Fluid antenna (FA) technology \cite{Wong2021Fluid} permits the radiating or receiving position to switch among multiple candidate ports within a compact aperture. FA implementations include mechanically movable antennas \cite{Zhu2024Movable} and liquid radiators controlled by micropumps or electrowetting-on-dielectric (EWOD) techniques \cite{Martinez2022Toward}. Pixelated reconfigurable antennas instead use positive-intrinsic-negative (PIN) diodes, micro-electro-mechanical-system (MEMS) switches, or related electronic switching networks to change the active radiating position without mechanically moving the antenna structure \cite{Zhang2025PixelAntenna,Wong2026Reconfigurable}. Electronic switching can support faster port reconfiguration, while hybrid architectures offer different tradeoffs between switching speed and hardware complexity \cite{New2026Fluid}.


Port selection in FAS has been studied for multiuser communications \cite{Wong2022FAMA}, integrated sensing and communication \cite{Ye2025FluidISAC}, and near-field transmission \cite{Chen2025JointBF_NF_FAS}. FA ports can also convey information through their activation indices. Existing FA-assisted IM schemes consider single-port activation \cite{Yang2024FA_PIM}, multiple-port activation \cite{Zhu2024FAIM}, correlation-aware grouping and coding \cite{Guo2025FAGIM,Faddoul2025CFAIM}, and differential transmission \cite{Zhang2026FA_RDIM}. In these schemes, the activated FA port or port pattern directly carries the index bits.

Reconfigurable intelligent surfaces (RISs) provide passive control of the reflected propagation channel. In RIS-aided receiver spatial modulation (RSM) and receiver spatial shift keying (RSSK), information bits are conveyed by receive-antenna indices, while the RIS phase shifts are adjusted to focus the reflected signal on the selected receive antenna \cite{Basar2020RIS_IM}. Extensions include joint transmit-and-receive indexing \cite{Ma2020LIS_SM_AS}, quadrature signaling \cite{Yuan2021RIS_RQRM,Dinan2022RIS_RQSSK}, generalized receive-antenna activation \cite{Marin2024RIS_RGSSK_RGSM}, and flexible receive-index mappings \cite{Yue2024IRS_EVSM,Dogukan2025RIS_ERSM,Yue2025RIS_FRSM}. Since these schemes assume fixed reflecting elements, their BER analyses do not include fluid-port selection or spatial correlation among candidate ports.

A fluid reconfigurable intelligent surface (FRIS) combines the phase reconfigurability of RIS with the position reconfigurability of FA \cite{Salem2025FRIS_FirstLook}. Existing FRIS studies have investigated rate and capacity \cite{VegaSanchez2026FRIS_PhasePosition,Rostami2025FRIS_Performance}, beamforming \cite{Xiao2026FRIS_PatternCoDesign,Xiao2025FRIS_JointOnOff}, holographic transmission \cite{Sun2026FRIS_Holographic}, and physical-layer security \cite{Kaveh2026FRIS_PLS,VegaSanchez2026FRIS_PLS}. However, FRIS-based IM and its BER analysis under finite-resolution phase shifts have not been sufficiently studied. With continuous phase shifts, the phase of each active link can be compensated and port selection can be based on the cascaded-channel magnitude. With discrete phase shifts, the residual quantization error reduces the in-phase focusing component and produces a quadrature leakage component. The active port must therefore be selected according to its post-quantization in-phase component. Moreover, the candidate ports are spatially correlated, and each receive-antenna index may select a different port set. When two selected port sets overlap, the corresponding equivalent channels are statistically dependent. Therefore, the PEP analysis requires their joint post-selection statistics.

In this paper, we propose fluid-port-selected receive index modulation (FPS-RIM) for FRIS-assisted single-input multiple-output (SIMO) transmission. For each receive-antenna index, one port is selected from every FRIS group and its reflection phase is configured to focus the reflected signal on the corresponding receive antenna. FPS-RIM supports two transmission modes. In the index-only (IO) mode, information bits are conveyed only by the receive-antenna index. In the symbol-augmented (SA) mode, the receive-antenna index and a conventional modulation symbol jointly carry information. The BER analysis mainly considers discrete phase shifts; continuous phase shifts are used as an ideal benchmark.

The main contributions of this paper are summarized as follows:

\begin{itemize}
  \item[$\bullet$] We propose discrete-phase FPS-RIM with index-only (IO) and symbol-augmented (SA) modes. For each receive-index state, one fluid port is selected from each group and its finite-resolution phase shift is configured. The residual phase error produces an in-phase focusing component and a quadrature leakage component. Accordingly, the active port is selected by maximizing the post-quantization in-phase component rather than the unquantized cascaded-channel magnitude.

  \item[$\bullet$] We analyze the BER performance of FPS-RIM over spatially correlated FRIS cascaded channels. Conditional Rician distributions and Marcum-$Q$-function integrals are used to obtain the joint moments of the equivalent channels after fluid-port selection. Based on these moments, correlated Gaussian distance models and quadratic-form Laplace transforms are used to derive the unconditional pairwise error probability (UPEP) and a bit-weighted BER union bound. The derivation accounts for overlap between the port sets selected for the transmitted and competing receive-index states. The analysis also covers continuous phase shifts as an ideal benchmark.

  \item[$\bullet$] We formulate the maximum-likelihood (ML) detector and develop greedy detection (GD) and a low-complexity Top-$L$ list detector for both FPS-RIM modes. After energy ranking, the Top-$L$ detector evaluates the joint Euclidean-distance metric for only $L$ receive-index candidates, reducing the number of tested index-symbol hypotheses from $N_{\mathrm r}M$ to $LM$. Numerical results compare the analytical and simulated BER and examine the effects of phase resolution, the number of candidate ports, the number of groups, the group aperture, and the list size.
\end{itemize}

The rest of this paper is organized as follows. Section~\ref{section_system_model} presents the FRIS architecture, correlated cascaded-channel model, and FPS-RIM transmission framework. Section~\ref{section_configuration_detection} presents the discrete-phase port and phase configuration and the corresponding detectors, with continuous phase included as an ideal benchmark. Section~\ref{section_performance_analysis} derives the UPEP and BER expressions. Section~\ref{sec:numerical_results} presents numerical results, and Section~\ref{sec:conclusion} concludes the paper.

  \emph{Notation}: Bold lowercase and uppercase letters represent vectors and matrices, respectively. The sets $\mathbb C$, $\mathbb R$, and $\mathbb R_+$ denote the complex, real, and nonnegative-real domains, respectively; $\mathbf I_n$ and $\mathbf 0_n$ are the $n\times n$ identity and zero matrices. The imaginary unit is $\mathrm{\jmath}=\sqrt{-1}$. For a complex scalar $x$, $|x|$ and $\angle(x)$ denote its magnitude and phase, while $\Re\{\cdot\}$ and $\Im\{\cdot\}$ extract the real and imaginary parts; the subscripts $\mathrm R$ and $\mathrm I$ identify real and imaginary components, respectively. The operators $(\cdot)^{\mathrm T}$ and $(\cdot)^{\mathrm H}$ denote transpose and conjugate transpose, $\|\cdot\|_2$ denotes the Euclidean norm, and $\det(\cdot)$ denotes the determinant. The floor operator is $\lfloor\cdot\rfloor$, and $(x)_{2\pi}$ denotes reduction modulo $2\pi$. For a random variable $X$, $f_X(\cdot)$ and $F_X(\cdot)$ denote its probability density function and cumulative distribution function, while $\mathbb E[\cdot]$ and $\Pr(\cdot)$ denote expectation and probability; $\mathbf 1\{\cdot\}$ is the indicator function. A circularly symmetric complex Gaussian random variable follows $\mathcal{CN}(\mu,\sigma^2)$, and a real Gaussian random vector follows $\mathcal N(\boldsymbol\mu,\mathbf C)$. The functions $Q(\cdot)$ and $Q_1(\cdot,\cdot)$ denote the Gaussian $Q$-function and first-order Marcum-$Q$ function, respectively; $J_0(\cdot)$ denotes the zeroth-order Bessel function of the first kind, and $I_0(\cdot)$ denotes the zeroth-order modified Bessel function of the first kind. Calligraphic symbols denote sets or alphabets.

  \section{System and Channel Model}
  \label{section_system_model}
  
\begin{figure}[!t]
    \centering
    \includegraphics[width=3.4in]{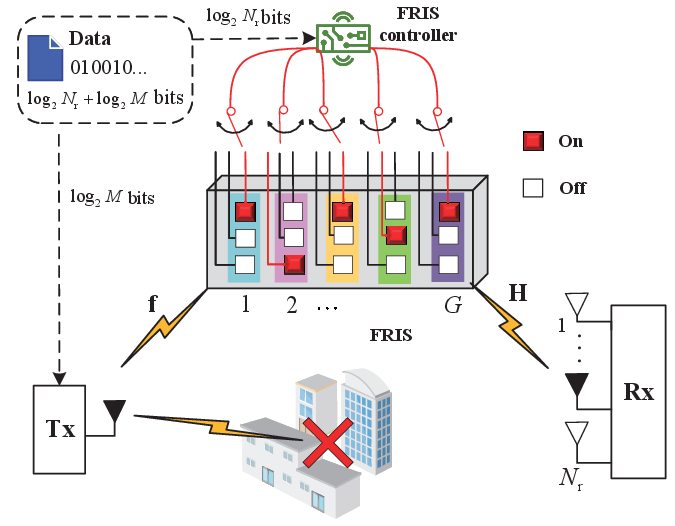}
    \caption{System model of the proposed FPS-RIM framework. \label{fig:system_model}}
  \end{figure}

  As illustrated in Fig.~\ref{fig:system_model}, we consider a FRIS-assisted single-input multiple-output (SIMO) system, where a single-antenna transmitter (Tx) communicates with an $N_{\text r}$-antenna receiver (Rx) through a FRIS. The direct Tx--Rx link is assumed to be blocked by surrounding obstacles. For each transmitted receive-index state, the FRIS controller selects the active fluid ports and configures their reflection phases to focus the reflected signal on the intended receive antenna.
  
  \subsection{FRIS Architecture and Spatial Correlation Channel Model}
  
  The FRIS comprises $N$ candidate fluid ports partitioned into $G$ groups, each containing $P$ ports, such that $N=GP$. During each channel use, one port is activated in each group, resulting in $G$ parallel fluid-port selections over the corresponding candidate-port sets. Each active port is connected to the centralized FRIS controller through a switching and phase control network \cite{Zhang2025PixelAntenna}.

  Each group occupies a one-dimensional aperture of length $W_{\mathrm g}\lambda$, where $\lambda$ is the carrier wavelength. The $P$ candidate ports are uniformly distributed over this aperture, and the position of the $p$-th port relative to the reference port is given by
  \begin{equation}
  d_p
  =
  \begin{cases}
  0, & P=1,\\
  \dfrac{(p-1)W_{\mathrm g}\lambda}{P-1}, & P>1,
  \end{cases}
  \qquad p=1,\ldots,P.
  \label{eq:group_port_position}
  \end{equation}
  The first port is used as the reference port of each group. The distance between any two ports belonging to different groups is designed to exceed half a wavelength, and the resulting inter-group spatial correlation is assumed negligible. In contrast, the more densely deployed candidate ports within each group are modeled as spatially correlated.

  Let $\mathbf f_g=[f_{g,1},\ldots,f_{g,P}]^{\mathrm T}\in\mathbb{C}^{P\times1}$ denote the Tx--FRIS channel vector of group $g$, where $f_{g,p}$ is the channel coefficient from the Tx to its $p$-th candidate port. Following the reference-port correlation model, the group-wise channel coefficients are expressed as
  \begin{equation}
  f_{g,p}
  =
  \rho_p f_{g,1}
  +
  \sqrt{1-|\rho_p|^2}\,\widetilde f_{g,p},
  \qquad p=2,\ldots,P,
  \label{eq:tx_fris_reference_channel}
  \end{equation}
  where $f_{g,1}$ and $\widetilde f_{g,p}$ are mutually independent $\mathcal{CN}(0,1)$ random variables. Following the Jakes model for isotropic scattering \cite{New2024Information}, the correlation coefficient between the $p$-th port and the reference port is
  \begin{equation}
  \rho_p
  =
  J_0\!\left(\frac{2\pi d_p}{\lambda}\right),
  \label{eq:reference_port_correlation}
  \end{equation}
  where $J_0(\cdot)$ denotes the Bessel function of the first kind of order zero.

  For the FRIS--Rx link, let $\mathbf H_g=[\mathbf h_{g,1},\ldots,\mathbf h_{g,P}]\in\mathbb{C}^{N_{\mathrm r}\times P}$ denote the channel matrix of group $g$, where $\mathbf h_{g,p}=[h_{1,g,p},\ldots,h_{N_{\mathrm r},g,p}]^{\mathrm T}$ is the channel vector from its $p$-th candidate port to the Rx. For receive antenna $r$, the channel coefficients within group $g$ are modeled as
  \begin{equation}
  h_{r,g,p}
  =
  \rho_p h_{r,g,1}
  +
  \sqrt{1-|\rho_p|^2}\,\widetilde h_{r,g,p},
  \qquad p=2,\ldots,P,
  \label{eq:fris_rx_reference_channel}
  \end{equation}
  where $h_{r,g,1}$ and $\widetilde h_{r,g,p}$ are mutually independent $\mathcal{CN}(0,1)$ random variables. The Tx--FRIS and FRIS--Rx links are mutually independent. Moreover, the FRIS--Rx channels associated with different receive antennas and different groups are independently generated.

  The cascaded channel through the $p$-th port of group $g$ toward receive antenna $r$ is defined as
  \begin{equation}
  c_{r,g,p}
  \triangleq
  h_{r,g,p}f_{g,p}.
  \label{eq:grouped_cascaded_channel}
  \end{equation}
  
  \subsection{FPS-RIM Transmission Framework}

  In the proposed fluid-port-selected receive index modulation (FPS-RIM) framework, the $N_{\mathrm r}$ receive antennas define $N_{\mathrm r}$ receive-index states. A receive-index state $i\in\{1,\ldots,N_{\mathrm r}\}$ specifies the intended receive antenna. For each state, the FRIS selects one fluid port from every group and configures its reflection phase to concentrate the reflected signal energy at that antenna. The selected receive-antenna index conveys the index bits.

  FPS-RIM supports two operating modes, namely the index-only (IO) mode and the symbol-augmented (SA) mode. Let $\mathbf b\in\{0,1\}^{B\times1}$ denote the input bit block. In FPS-RIM-IO, all bits are mapped to the receive-index state $i$, and no additional modulation symbol is transmitted. In FPS-RIM-SA, the receive-index bits select $i$, while an additional $\log_2M$ bits select a unit-average-energy symbol $s_m\in\mathcal{X}$ from an $M$-ary phase-shift keying (PSK) or quadrature amplitude modulation (QAM) constellation. The number of transmitted bits per channel use is therefore
  \begin{equation}
  B
  =
  \begin{cases}
  \log_2N_{\mathrm r}, & \text{FPS-RIM-IO},\\
  \log_2N_{\mathrm r}+\log_2M, & \text{FPS-RIM-SA}.
  \end{cases}
  \label{eq:fps_rim_bits}
  \end{equation}
  Both modes can be represented by the transmit state $(i,s_m)$ by setting $\mathcal{X}=\{1\}$ and $M=1$ for FPS-RIM-IO. Accordingly, the IO and SA modes contain $N_{\mathrm r}$ and $N_{\mathrm r}M$ candidate transmit states, respectively.

  For receive-index state $i$, let $p_{i,g}^{\star}\in\{1,\ldots,P\}$ denote the active port selected from group $g$, and let $\theta_{i,g}^{\star}\in[0,2\pi)$ denote its reflection phase. The state-dependent port and phase configurations are collected in $\mathbf p_i^{\star}=[p_{i,1}^{\star},\ldots,p_{i,G}^{\star}]^{\mathrm T}$ and $\boldsymbol{\theta}_i^{\star}=[\theta_{i,1}^{\star},\ldots,\theta_{i,G}^{\star}]^{\mathrm T}$, respectively. Since the FRIS is configured separately for each intended antenna, two receive-index states may select identical, partially overlapping, or distinct sets of fluid ports.

  For receive-index state $i$, the effective cascaded channel at receive antenna $r$ is
  \begin{equation}
  H_r^{(i)}
  =
  \sum_{g=1}^{G}
  c_{r,g,p_{i,g}^{\star}}
  e^{\mathrm{\jmath}\theta_{i,g}^{\star}}.
  \label{eq:H_eff_def}
  \end{equation}
  Collecting these gains gives the equivalent channel vector associated with receive-index state $i$ as $\mathbf z_i=[H_1^{(i)},\ldots,H_{N_{\mathrm r}}^{(i)}]^{\mathrm T}$. The received signal corresponding to transmit state $(i,s_m)$ is then expressed as
  \begin{equation}
  \mathbf y
  =
  \sqrt{E_s}\,\mathbf z_i s_m
  +
  \mathbf n,
  \label{eq:fps_rim_received_signal}
  \end{equation}
  where $E_s$ is the energy per channel use, $\mathbb{E}[|s_m|^2]=1$, and $\mathbf n\sim\mathcal{CN}(\mathbf 0,N_0\mathbf I_{N_{\mathrm r}})$. The instantaneous signal-to-noise ratio (SNR) at receive antenna $r$ is
  \begin{equation}
  \gamma_r^{(i)}
  =
  \frac{E_s}{N_0}\big|H_r^{(i)}\big|^2.
  \label{eq:SNR_r_def}
  \end{equation}
  In FPS-RIM-IO, $s_m=1$ and the received signal is determined solely by the receive-index state. FPS-RIM-SA retains the same grouped port-selection and focusing mechanism while jointly conveying the constellation symbol $s_m$. 
  
  \section{FPS-RIM Configuration and Detection}
\label{section_configuration_detection}

  The FRIS configuration associated with receive-index state $i$ is determined independently in each port group and is common to the IO and SA modes. In particular, the selected port and reflection phase depend on the cascaded channels toward the intended antenna but not on the constellation symbol $s_m$.

  \subsection{Discrete-Phase Port and Phase Configuration}

  We consider $Q$-bit phase control with phase spacing $\Delta_Q=2\pi/2^Q$. The available phase alphabet is
  \begin{equation}
  \mathcal{A}_Q
  =
  \left\{0,\Delta_Q,2\Delta_Q,\ldots,(2^Q-1)\Delta_Q\right\}.
  \label{eq:phase_alphabet}
  \end{equation}
  For each candidate port, the ideal compensation phase $\big(-\angle(c_{i,g,p})\big)_{2\pi}$ is mapped to its nearest element in $\mathcal{A}_Q$ according to
  \begin{equation}
  \theta_{i,g,p}^{(Q)}
  =
  \left(
  \Delta_Q
  \left\lfloor
  \frac{\big(-\angle(c_{i,g,p})\big)_{2\pi}}{\Delta_Q}
  +
  \frac{1}{2}
  \right\rfloor
  \right)_{2\pi}.
  \label{eq:nearest_phase_quantizer}
  \end{equation}
  The inner modulo operation maps the desired phase to $[0,2\pi)$, whereas the outer modulo operation identifies $2\pi$ with zero after nearest-neighbor quantization. Consequently, the quantizer is circular across the phase boundary, and the resulting wrapped phase error satisfies $|\varepsilon_{i,g,p}|\leq\Delta_Q/2$.

  Finite-resolution phase control generally leaves a residual phase error. The useful in-phase contribution produced by candidate port $p$ in group $g$ is therefore $\Re\!\left\{
  c_{i,g,p}e^{\mathrm{\jmath}\theta_{i,g,p}^{(Q)}}
  \right\}$.
  Maximizing this post-quantization in-phase projection yields the selected port index and corresponding reflection phase as
  \begin{equation}
  p_{i,g}^{\star}
  =
  \arg\max_{1\le p\le P}\Re\!\left\{
  c_{i,g,p}e^{\mathrm{\jmath}\theta_{i,g,p}^{(Q)}}\right\},
  \qquad
  \theta_{i,g}^{\star}
  =
  \theta_{i,g,p_{i,g}^{\star}}^{(Q)}.
  \label{eq:quantized_group_selection}
  \end{equation}
  The resulting effective channel at the intended antenna is
  \begin{equation}
  H_i^{(i)}
  =
  \sum_{g=1}^{G}
  c_{i,g,p_{i,g}^{\star}}
  e^{\mathrm{\jmath}\theta_{i,g}^{\star}}.
  \label{eq:quantized_focused_gain}
  \end{equation}
  
  \subsection{Continuous-Phase Benchmark}

  With continuous phase control, the candidate phase for port $p$ in group $g$ is chosen to compensate for the phase of its cascaded channel toward the intended antenna:
  \begin{equation}
  \theta_{i,g,p}^{(\infty)}
  =
  \big(-\angle(c_{i,g,p})\big)_{2\pi}.
  \label{eq:continuous_candidate_phase}
  \end{equation}
  This choice gives $c_{i,g,p}e^{\mathrm{\jmath}\theta_{i,g,p}^{(\infty)}}=|c_{i,g,p}|$. Since exactly one port is activated in each group, maximizing the focused contribution of group $g$ reduces to
  \begin{equation}
  p_{i,g}^{\star}
  =
  \arg\max_{1\le p\le P}|c_{i,g,p}|,
  \qquad
  \theta_{i,g}^{\star}
  =
  \theta_{i,g,p_{i,g}^{\star}}^{(\infty)}.
  \label{eq:continuous_group_selection}
  \end{equation}
  Accordingly, the effective channel at the intended antenna is real and nonnegative and can be written as
  \begin{equation}
  H_i^{(i)}
  =
  \sum_{g=1}^{G}|c_{i,g,p_{i,g}^{\star}}|
  =
  \sum_{g=1}^{G}|h_{i,g,p_{i,g}^{\star}}|\,|f_{g,p_{i,g}^{\star}}|.
  \label{eq:continuous_focused_gain}
  \end{equation}

  \subsection{Conditional Distribution and Detection}

  Define the equivalent channel matrix as
  $
  \mathbf Z
  =
  \left[\mathbf z_1,\mathbf z_2,\ldots,\mathbf z_{N_{\mathrm r}}\right]
  \in\mathbb{C}^{N_{\mathrm r}\times N_{\mathrm r}},
  $
  where column $\mathbf z_i$ incorporates the grouped port selection and phase configuration associated with receive-index state $i$. We assume that $\mathbf Z$ is known at the receiver. Since two receive-index states may select the same port in some groups, the columns of $\mathbf Z$ are generally statistically dependent. The detectors developed below operate directly on their realized channel vectors and do not require an independence assumption.

  \subsubsection{Conditional Probability Density Function (PDF)}

  Conditioned on $\mathbf Z$ and transmit state $(i,s_m)$, the received vector in \eqref{eq:fps_rim_received_signal} follows $\mathbf y\mid(\mathbf Z,i,s_m)\sim\mathcal{CN}\!\left(\sqrt{E_s}\mathbf z_i s_m,N_0\mathbf I_{N_{\mathrm r}}\right)$.
  Its conditional probability density function is therefore
  \begin{equation}
  p\!\left(\mathbf y\mid\mathbf Z,i,s_m\right)
  =
  \frac{1}{(\pi N_0)^{N_{\mathrm r}}}
  \exp\!\left(
  -\frac{\left\|\mathbf y-\sqrt{E_s}\mathbf z_i s_m\right\|_2^2}{N_0}
  \right).
  \label{eq:y_cond_pdf_cascaded}
  \end{equation}
  For the IO mode, $M=1$ and $s_1=1$, so the conditional distribution is determined solely by receive-index state $i$. The SA mode requires the receiver to recover both the receive-index state and the constellation symbol.

  \subsubsection{ML Detection}

  The maximum-likelihood (ML) detector jointly searches all candidate transmit states according to
  \begin{equation}
  (\widehat i_{\mathrm{ML}},\widehat m_{\mathrm{ML}})
  =
  \arg\min_{\substack{1\le i\le N_{\mathrm r}\\1\le m\le M}}
  \left\|
  \mathbf y-\sqrt{E_s}\mathbf z_i s_m
  \right\|_2^2.
  \label{eq:ML_joint_cascaded}
  \end{equation}
  The SA mode requires $N_{\mathrm r}M$ full-vector distance evaluations. For the IO mode, setting $M=1$ and $s_1=1$ reduces the search to the $N_{\mathrm r}$ receive-index states. Since ML detection uses the amplitude and phase observations from all receive antennas, it serves as the performance benchmark for the reduced-complexity detectors.

  \subsubsection{Greedy Detection}

  The FRIS focusing operation generally produces a power disparity between the intended and non-intended receive antennas. The greedy detector (GD) exploits this property and first estimates the receive-index state as
  \begin{equation}
  \widehat i_{\mathrm G}
  =
  \arg\max_{1\le r\le N_{\mathrm r}}|y_r|^2.
  \label{eq:greedy_mode}
  \end{equation}
  For the IO mode, this decision directly determines the detected transmit state. For the SA mode, the constellation symbol is subsequently detected using the complete received vector:
  \begin{equation}
  \widehat m_{\mathrm G}
  =
  \arg\min_{1\le m\le M}
  \left\|
  \mathbf y-\sqrt{E_s}\mathbf z_{\widehat i_{\mathrm G}}s_m
  \right\|_2^2.
  \label{eq:greedy_symbol}
  \end{equation}
  Thus, symbol detection is not performed from only the scalar observation at antenna $\widehat i_{\mathrm G}$; it retains all $N_{\mathrm r}$ receive observations for the selected receive-index state. GD evaluates $M$ full-vector distances in the SA mode and requires no additional constellation search in the IO mode. The accuracy of the first-stage decision depends on the power difference between the intended and non-intended antennas; an incorrect index decision also propagates to symbol detection.

  \subsubsection{Top-\texorpdfstring{$L$}{L} List Detection}

  The Top-$L$ detector retains the $L$ receive-index states with the largest received energies before evaluating the joint distance metric. Let $\pi(\cdot)$ be a permutation of $\{1,\ldots,N_{\mathrm r}\}$ satisfying
  \begin{equation}
  |y_{\pi(1)}|^2
  \ge
  |y_{\pi(2)}|^2
  \ge
  \cdots
  \ge
  |y_{\pi(N_{\mathrm r})}|^2,
  \label{eq:receive_energy_order}
  \end{equation}
  and define the Top-$L$ receive-index candidate set as $\mathcal I_L=\{\pi(1),\ldots,\pi(L)\}$, where $1\le L\le N_{\mathrm r}$.
  The second stage restricts the joint ML search to the transmit states associated with $\mathcal I_L$:
  \begin{equation}
  (\widehat i_L,\widehat m_L)
  =
  \arg\min_{\substack{i\in\mathcal I_L\\1\le m\le M}}
  \left\|
  \mathbf y-\sqrt{E_s}\mathbf z_i s_m
  \right\|_2^2.
  \label{eq:list_nearML_joint}
  \end{equation}
  Computing one full-vector Euclidean-distance metric requires $\mathcal O(N_{\mathrm r})$ operations. Exhaustive ML detection evaluates $N_{\mathrm r}M$ such metrics and therefore has complexity $\mathcal O(N_{\mathrm r}^2M)$. The Top-$L$ detector requires $\mathcal O(N_{\mathrm r}\log N_{\mathrm r})$ operations to rank the received energies and $\mathcal O(LMN_{\mathrm r})$ operations to evaluate the retained hypotheses, giving an overall complexity of $\mathcal O(N_{\mathrm r}\log N_{\mathrm r}+LMN_{\mathrm r})$. Thus, for $L\ll N_{\mathrm r}$, its distance-search complexity is approximately $L/N_{\mathrm r}$ of that required by ML. At $L=1$, a direct maximum search avoids full sorting, resulting in $\mathcal O(N_{\mathrm r})$ complexity for FPS-RIM-IO and $\mathcal O(MN_{\mathrm r})$ for FPS-RIM-SA. At $L=N_{\mathrm r}$, candidate pruning is absent and \eqref{eq:list_nearML_joint} is identical to ML detection.

\section{Performance Analysis}
\label{section_performance_analysis}

This section derives the BER of ML detection for FPS-RIM-IO and FPS-RIM-SA. The finite-resolution analysis begins with the post-selection statistics of one spatially correlated port group, including the dependence between phase quantization and port selection. The resulting moments specify the joint Gaussian distance models for receive-index and symbol errors, from which the UPEP and BER union bound follow. Continuous phase control is treated as the ideal-phase benchmark.

\subsection{Conditional PEP}

Consider the transmit state $(i,s_m)$ and a competing state $(\widehat i,s_{\widehat m})$. Conditioned on the equivalent channel matrix $\mathbf Z$, the pairwise error probability (PEP) is
\begin{equation}
P\!\left((i,m)\rightarrow(\widehat i,\widehat m)\mid\mathbf Z\right)
=Q\!\left(\sqrt{\frac{\gamma}{2}\mathcal D_{i,\widehat i}^{m,\widehat m}}\right),
\label{eq:new_cpep}
\end{equation}
where $\gamma=E_s/N_0$ and
\begin{equation}
\mathcal D_{i,\widehat i}^{m,\widehat m}
=\left\|s_m\mathbf z_i-s_{\widehat m}\mathbf z_{\widehat i}\right\|_2^2.
\label{eq:new_random_distance}
\end{equation}
Define the Laplace transform of the random distance as
\begin{equation}
\mathcal L_{\mathcal D}(t)
=\int_0^\infty e^{-t\delta}f_{\mathcal D}(\delta)\,d\delta.
\label{eq:new_distance_laplace}
\end{equation}
Craig's representation of the Gaussian $Q$-function gives the average PEP
\begin{equation}
\overline P_{i,\widehat i}^{m,\widehat m}
=\frac{1}{\pi}\int_0^{\pi/2}
\mathcal L_{\mathcal D}\!\left(\frac{\gamma}{4\sin^2\vartheta}\right)d\vartheta.
\label{eq:new_craig_pep}
\end{equation}
Using the classical approximation $Q(x)\simeq\frac{1}{12}\exp(-x^2/2)+\frac{1}{4}\exp(-2x^2/3)$, a computationally convenient form is
\begin{equation}
\overline P_{i,\widehat i}^{m,\widehat m}
\simeq
\frac{1}{12}\mathcal L_{\mathcal D}\!\left(\frac{\gamma}{4}\right)
+\frac{1}{4}\mathcal L_{\mathcal D}\!\left(\frac{\gamma}{3}\right).
\label{eq:new_two_exp_pep}
\end{equation}

\subsection{Selection-Aware Port Statistics}

The group index is omitted in this subsection. Define the Tx--FRIS and FRIS--Rx port powers as $X_p=|f_p|^2$ and $Y_{i,p}=|h_{i,p}|^2$, respectively. For $p\ge2$, conditioning on the reference-port power $Y_{i,1}=u$ gives
\begin{equation}
\begin{aligned}
f_{Y_{i,p}|Y_{i,1}}(y|u)
&=\frac{1}{1-|\rho_p|^2}
\exp\!\left(-\frac{y+|\rho_p|^2u}{1-|\rho_p|^2}\right)
\\[-1mm]
&\quad{}\times I_0\!\left(\frac{2\sqrt{|\rho_p|^2uy}}{1-|\rho_p|^2}\right).
\end{aligned}
\label{eq:new_conditional_rx_power_pdf}
\end{equation}
The first two conditional moments following from \eqref{eq:new_conditional_rx_power_pdf} are
\begin{equation}
\begin{aligned}
\mathbb E[Y_{i,p}\mid u]
&=1-|\rho_p|^2+|\rho_p|^2u,\\
\mathbb E[Y_{i,p}^2\mid u]
&=2(1-|\rho_p|^2)^2
+4|\rho_p|^2(1-|\rho_p|^2)u
+|\rho_p|^4u^2.
\end{aligned}
\label{eq:new_conditional_rx_power_moments}
\end{equation}
The conditional density and moments of $X_p|X_1$ have the same forms. These explicit moments will be used to average channel coefficients that do not participate in the port-selection event. Hence, the joint PDF of the source-link power vector $\mathbf x=[x_1,\ldots,x_P]^{\mathrm T}$ is
\begin{equation}
f_{\mathbf X}(\mathbf x)
=e^{-x_1}\prod_{p=2}^P f_{X_p|X_1}(x_p|x_1),
\qquad \mathbf x\in\mathbb R_+^P.
\label{eq:new_source_power_joint_pdf}
\end{equation}
All subsequent integrations are evaluated with respect to these conditional Rician-power distributions.

\subsubsection{Discrete-Phase Port Selection}

\begin{figure}[!t]
\centering
\includegraphics[width=0.85\columnwidth]{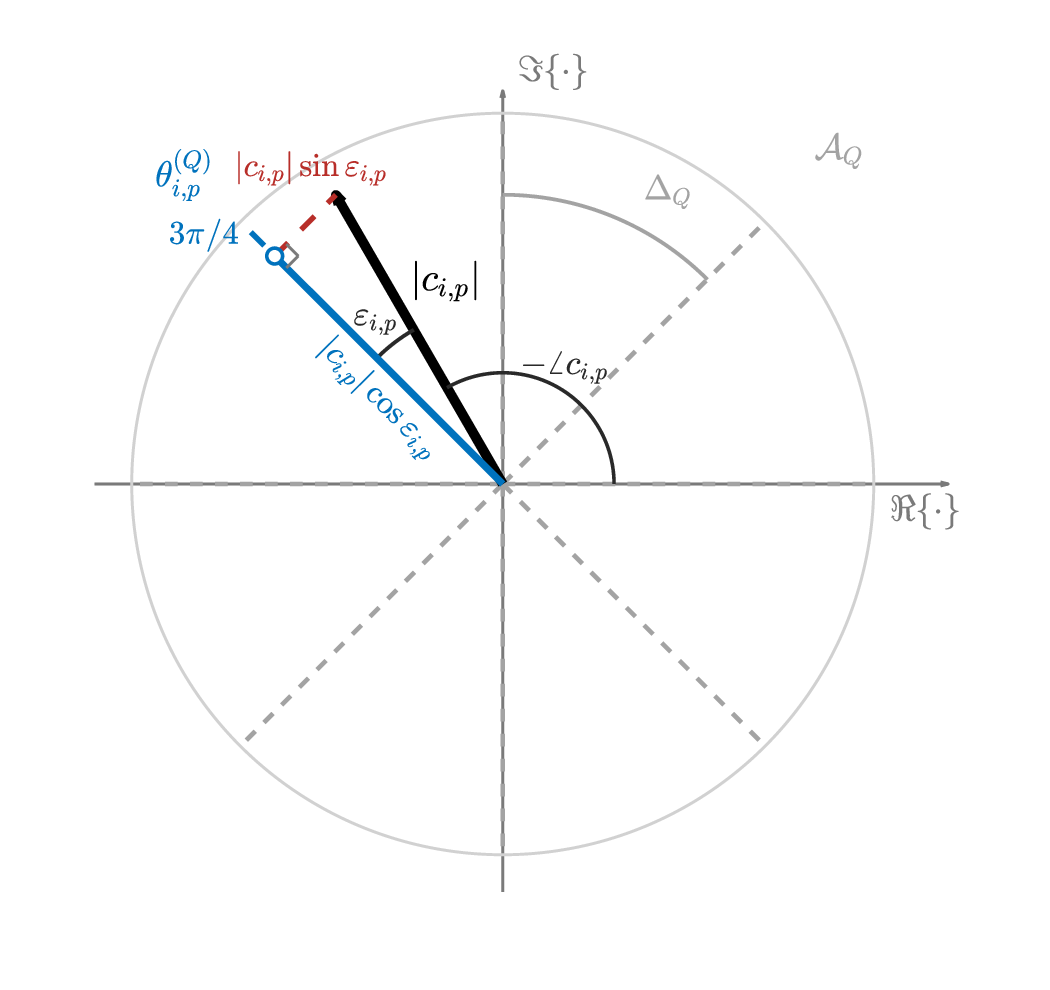}
\caption{Finite-resolution phase alignment and decomposition of the cascaded-channel contribution into its in-phase focusing and quadrature leakage components.}
\label{fig:quantized_phase_projection_geometry}
\end{figure}

For $Q$-bit phase control, let $\varepsilon_{i,p}\in[-\Delta_Q/2,\Delta_Q/2]$ denote the residual phase error. The phase-rotated cascaded contribution and its in-phase selection score are
\begin{equation}
V_{i,p}^{\mathrm Q}
=\sqrt{X_pY_{i,p}}e^{\mathrm j\varepsilon_{i,p}},
\qquad
S_{i,p}^{\mathrm Q}
=\sqrt{X_pY_{i,p}}\cos\varepsilon_{i,p},
\label{eq:new_quantized_contribution_score}
\end{equation}
Fig.~\ref{fig:quantized_phase_projection_geometry} depicts finite-resolution phase alignment in the compensation-phase plane. The ideal compensation phase $-\angle c_{i,p}$ is mapped to the nearest $\theta_{i,p}^{(Q)}\in\mathcal A_Q$, leaving the wrapped error $\varepsilon_{i,p}$. The resulting cascaded-channel contribution is decomposed into the in-phase projection $|c_{i,p}|\cos\varepsilon_{i,p}$, which forms the port-selection score, and the orthogonal residual $|c_{i,p}|\sin\varepsilon_{i,p}$, which represents phase-quantization-induced quadrature leakage.
Accordingly, $p_i^\star=\arg\max_{1\le p\le P}S_{i,p}^{\mathrm Q}$. Let $R_{i,p}=\sqrt{xY_{i,p}}$ and $a_Q=\Delta_Q/2$. Given $X_p=x$ and $Y_{i,1}=u$, its conditional density is $f_{R_{i,p}}(r|x,u)=2r f_{Y_{i,p}|Y_{i,1}}(r^2/x|u)/x$. Averaging the uniform phase error analytically gives the conditional score CDF in \eqref{eq:new_quantized_score_cdf} for $s\ge0$.
\begin{figure*}[!b]
\centering
\rule{\textwidth}{0.4pt}
\vspace{-1.5ex}
\begin{equation}
F_{S_{i,p}^{\mathrm Q}}(s|x,u)=F_{R_{i,p}}(s|x,u)+\int_s^{s/\cos a_Q}\left[1-\frac{\arccos(s/r)}{a_Q}\right]f_{R_{i,p}}(r|x,u)\,dr,
\label{eq:new_quantized_score_cdf}
\end{equation}
\end{figure*}
Define the quantized-phase selected-port kernel as
\begin{equation}
\begin{aligned}
w_p^{\mathrm Q}(\mathbf x,u,y,\varepsilon)
&=\frac{f_{\mathbf X}(\mathbf x)e^{-u}}{\Delta_Q}
f_{Y_{i,p}|Y_{i,1}}(y|u)\\
&\quad{}\times
\prod_{q\ne p}F_{S_{i,q}^{\mathrm Q}}\!\left(
\sqrt{x_py}\cos\varepsilon\mid x_q,u
\right).
\end{aligned}
\label{eq:new_quantized_selection_kernel}
\end{equation}
The required trigonometric moments are
\begin{equation}
\begin{aligned}
M_{a,b}^{\mathrm Q}
&=\sum_{p=1}^P
\int_{\mathbb R_+^P}\int_0^\infty\int_0^\infty
\int_{-\Delta_Q/2}^{\Delta_Q/2}
(\sqrt{x_py}\cos\varepsilon)^a\\
&\quad{}\times(\sqrt{x_py}\sin\varepsilon)^b
w_p^{\mathrm Q}(\mathbf x,u,y,\varepsilon)
\,d\varepsilon\,dy\,du\,d\mathbf x,
\end{aligned}
\label{eq:new_quantized_trigonometric_moments}
\end{equation}
where the pairs $(a,b)$ required below satisfy $a+b\le2$. The phase-error integrations in \eqref{eq:new_quantized_trigonometric_moments} can be evaluated explicitly. In particular, let $t$ be the largest competing score and let $r=\sqrt{x_py}$. For $r>t$, port $p$ is selected only over the symmetric phase interval $|\varepsilon|\le\phi(r,t)$, where
\begin{equation}
\phi(r,t)=\min\!\left\{\frac{\Delta_Q}{2},\arccos\!\left(\frac{t}{r}\right)\right\}.
\label{eq:new_admissible_phase_halfwidth}
\end{equation}
Over this interval, the required normalized phase integrals are
\begin{equation}
\begin{aligned}
\frac{1}{\Delta_Q}\int_{-\phi}^{\phi}d\varepsilon
&=\frac{2\phi}{\Delta_Q},\\
\frac{1}{\Delta_Q}\int_{-\phi}^{\phi}\cos\varepsilon\,d\varepsilon
&=\frac{2\sin\phi}{\Delta_Q},\\
\frac{1}{\Delta_Q}\int_{-\phi}^{\phi}\cos^2\varepsilon\,d\varepsilon
&=\frac{\phi+\frac{1}{2}\sin(2\phi)}{\Delta_Q},\\
\frac{1}{\Delta_Q}\int_{-\phi}^{\phi}\sin^2\varepsilon\,d\varepsilon
&=\frac{\phi-\frac{1}{2}\sin(2\phi)}{\Delta_Q}.
\end{aligned}
\label{eq:new_selected_phase_integrals}
\end{equation}
All integrals containing an odd power of $\sin\varepsilon$ vanish by symmetry. Thus, the phase variable can be integrated analytically after conditioning on $(r,t)$, leaving only the fading and competing-score averages in \eqref{eq:new_quantized_trigonometric_moments}. The focused two-dimensional mean and covariance are given by \eqref{eq:new_quantized_focused_mean_cov_general}.
\begin{figure*}[!b]
\centering
\begin{equation}
\boldsymbol\mu_f^{\mathrm Q}
=
\begin{bmatrix}M_{1,0}^{\mathrm Q}\\M_{0,1}^{\mathrm Q}\end{bmatrix},
\qquad
\mathbf C_f^{\mathrm Q}
=
\begin{bmatrix}
M_{2,0}^{\mathrm Q}-(M_{1,0}^{\mathrm Q})^2&M_{1,1}^{\mathrm Q}-M_{1,0}^{\mathrm Q}M_{0,1}^{\mathrm Q}\\
M_{1,1}^{\mathrm Q}-M_{1,0}^{\mathrm Q}M_{0,1}^{\mathrm Q}&M_{0,2}^{\mathrm Q}-(M_{0,1}^{\mathrm Q})^2
\end{bmatrix}.
\label{eq:new_quantized_focused_mean_cov_general}
\end{equation}
\end{figure*}
For the symmetric nearest-neighbor phase quantizer, $M_{0,1}^{\mathrm Q}=M_{1,1}^{\mathrm Q}=0$, and hence
\begin{equation}
\boldsymbol\mu_f^{\mathrm Q}
=\begin{bmatrix}M_{1,0}^{\mathrm Q}\\0\end{bmatrix},
\qquad
\mathbf C_f^{\mathrm Q}
=\begin{bmatrix}
M_{2,0}^{\mathrm Q}-(M_{1,0}^{\mathrm Q})^2&0\\
0&M_{0,2}^{\mathrm Q}
\end{bmatrix}.
\label{eq:new_quantized_focused_mean_cov_symmetric}
\end{equation}
Accordingly, $M_{1,0}^{\mathrm Q}$ represents the useful in-phase focusing amplitude, $M_{2,0}^{\mathrm Q}-(M_{1,0}^{\mathrm Q})^2$ describes its post-selection fluctuation, and $M_{0,2}^{\mathrm Q}$ is the residual quadrature energy caused by finite-resolution phase control. The last term vanishes under ideal continuous phase alignment.
To characterize the dependence between the focused gains of two receive-index states, define the conditional focused-mean vector directly by \eqref{eq:new_quantized_conditional_mean_vector}.
\begin{figure*}[!b]
\centering
\begin{equation}
\mathbf m_{\mathrm Q}(\mathbf x)=\sum_{p=1}^P\int_0^\infty\int_0^\infty e^{-u}f_{Y_{i,p}|Y_{i,1}}(y|u)\int_{-\Delta_Q/2}^{\Delta_Q/2}\frac{\sqrt{x_py}}{\Delta_Q}\begin{bmatrix}\cos\varepsilon\\\sin\varepsilon\end{bmatrix}\prod_{q\ne p}F_{S_{i,q}^{\mathrm Q}}\!\left(\sqrt{x_py}\cos\varepsilon\mid x_q,u\right)d\varepsilon\,dy\,du.
\label{eq:new_quantized_conditional_mean_vector}
\end{equation}
\end{figure*}
The focused cross-covariance matrix is then
\begin{equation}
\mathbf C_{fd}^{\mathrm Q}
=\int_{\mathbb R_+^P}f_{\mathbf X}(\mathbf x)
\mathbf m_{\mathrm Q}(\mathbf x)\mathbf m_{\mathrm Q}^{\mathrm T}(\mathbf x)d\mathbf x
-\boldsymbol\mu_f^{\mathrm Q}(\boldsymbol\mu_f^{\mathrm Q})^{\mathrm T}.
\label{eq:new_quantized_focused_cross_covariance}
\end{equation}
For the real reference-port coefficients in the adopted $J_0$ model and the symmetric nearest-neighbor quantizer, the second entry of $\mathbf m_{\mathrm Q}(\mathbf x)$ is zero. Hence,
\begin{equation}
\mathbf C_{fd}^{\mathrm Q}
=\begin{bmatrix}\xi_{\mathrm Q}&0\\0&0\end{bmatrix},
\quad
\xi_{\mathrm Q}=\int_{\mathbb R_+^P}f_{\mathbf X}(\mathbf x)m_{\mathrm Q,\mathrm R}^2(\mathbf x)d\mathbf x-(M_{1,0}^{\mathrm Q})^2.
\label{eq:new_quantized_focused_cross_covariance_symmetric}
\end{equation}
Thus, $\xi_{\mathrm Q}$ preserves the correlation between the useful in-phase focused components generated for two receive-index states, including the coupling introduced by their possibly overlapping selected-port sets; no quadrature cross-state covariance remains by symmetry.

\subsubsection{Continuous-Phase Benchmark}

Under continuous phase control, the focused amplitude provided by port $p$ is
\begin{equation}
R_{i,p}=\sqrt{X_pY_{i,p}},
\qquad
p_i^\star=\arg\max_{1\le p\le P}R_{i,p}.
\label{eq:new_continuous_selection_score}
\end{equation}
Given $X_p=x$ and $Y_{i,1}=u$, the conditional cumulative distribution function (CDF) ~\cite{Wong2021Fluid} of $R_{i,p}$ is
\begin{equation}
F_{R_{i,p}}(r|x,u)
=1-Q_1\!\left(
\sqrt{\frac{2|\rho_p|^2u}{1-|\rho_p|^2}},
\sqrt{\frac{2r^2}{(1-|\rho_p|^2)x}}
\right),
\label{eq:new_continuous_score_cdf}
\end{equation}
where the reference port is represented by its degenerate conditional distribution. The selected-port integration kernel is
\begin{equation}
\begin{aligned}
w_p^{\mathrm C}(\mathbf x,u,y)
&=f_{\mathbf X}(\mathbf x)e^{-u}f_{Y_{i,p}|Y_{i,1}}(y|u)\\
&\quad{}\times
\prod_{q\ne p}
F_{R_{i,q}}\!\left(\sqrt{x_py}\mid x_q,u\right).
\end{aligned}
\label{eq:new_continuous_selection_kernel}
\end{equation}
The $k$-th selected focused-amplitude moment, for $k\in\{1,2\}$, is therefore
\begin{equation}
M_k^{\mathrm C}
=\sum_{p=1}^P
\int_{\mathbb R_+^P}\int_0^\infty\int_0^\infty
(x_py)^{k/2}w_p^{\mathrm C}(\mathbf x,u,y)
\,dy\,du\,d\mathbf x.
\label{eq:new_continuous_selected_moments}
\end{equation}
Thus, $M_1^{\mathrm C}$ is the average coherent focusing amplitude contributed by one group, whereas the corresponding post-selection variance is
\begin{equation}
\sigma_{\mathrm C}^2=M_2^{\mathrm C}-(M_1^{\mathrm C})^2.
\label{eq:new_continuous_mean_variance}
\end{equation}
The variance $\sigma_{\mathrm C}^2$ captures both small-scale fading and the randomness introduced by selecting among correlated candidate ports.
The focused gains associated with two receive-index states remain dependent through the common Tx--FRIS powers. Their conditional focused mean is
\begin{equation}
\begin{aligned}
m_{\mathrm C}(\mathbf x)
&=\sum_{p=1}^P\int_0^\infty\int_0^\infty
\sqrt{x_py}\,e^{-u}f_{Y_{i,p}|Y_{i,1}}(y|u)\\
&\quad{}\times
\prod_{q\ne p}F_{R_{i,q}}\!\left(\sqrt{x_py}\mid x_q,u\right)
\,dy\,du.
\end{aligned}
\label{eq:new_continuous_conditional_mean}
\end{equation}
The corresponding focused cross covariance is
\begin{equation}
\xi_{\mathrm C}
=\int_{\mathbb R_+^P}f_{\mathbf X}(\mathbf x)m_{\mathrm C}^2(\mathbf x)d\mathbf x
-(M_1^{\mathrm C})^2.
\label{eq:new_continuous_focused_cross_covariance}
\end{equation}
Thus, $\xi_{\mathrm C}$ is not an independent leakage term; it quantifies the cross-state dependence between two coherently focused gains induced by the common Tx--FRIS channel and state-dependent port selection.

\subsection{Joint Gaussian Characterization}

For the group under consideration, define the contribution observed at receive antenna $r$ under state $i$ and phase-control case $b\in\{\mathrm C,\mathrm Q\}$ as
\begin{equation}
U_r^{i,b}
=h_{r,p_i^{\star,b}}f_{p_i^{\star,b}}
e^{\mathrm j\theta_i^{\star,b}}.
\label{eq:new_selected_group_contribution}
\end{equation}

\subsubsection{Discrete-Phase Configuration}

For quantized phase control, the per-group real-valued contribution vector is given by \eqref{eq:new_quantized_q}.
\begin{figure*}[!b]
\centering
\begin{equation}
\mathbf q^{\mathrm Q}
=\left[
\Re\{U_i^{i,\mathrm Q}\},\Im\{U_i^{i,\mathrm Q}\},
\Re\{U_i^{\widehat i,\mathrm Q}\},\Im\{U_i^{\widehat i,\mathrm Q}\},
\Re\{U_{\widehat i}^{i,\mathrm Q}\},\Im\{U_{\widehat i}^{i,\mathrm Q}\},
\Re\{U_{\widehat i}^{\widehat i,\mathrm Q}\},\Im\{U_{\widehat i}^{\widehat i,\mathrm Q}\}
\right]^{\mathrm T}.
\label{eq:new_quantized_q}
\end{equation}
\end{figure*}
Let $d\Lambda^{\mathrm Q}$ denote the joint selection measure generated by the common source-link vector, the two target-link vectors, their quantization errors, and the corresponding in-phase-projection selection events. Define
\begin{equation}
\eta_{\mathrm Q}=\int|U_i^{\widehat i,\mathrm Q}|^2d\Lambda^{\mathrm Q},
\qquad
\tau_{\mathrm Q}=\int U_i^{\widehat i,\mathrm Q}U_{\widehat i}^{i,\mathrm Q}d\Lambda^{\mathrm Q}.
\label{eq:new_quantized_eta_tau}
\end{equation}
Here, $\eta_{\mathrm Q}$ measures the off-target energy under quantized phase control, while $\tau_{\mathrm Q}$ retains the reciprocal coupling between the two off-target contributions after both phase quantization and port selection. Under the real reference-port model and symmetric quantization interval, conjugate phase-error realizations occur with equal probability, and therefore $\tau_{\mathrm Q}\in\mathbb R$.
Its mean is
\begin{equation}
\boldsymbol\mu_q^{\mathrm Q}
=\left[M_{1,0}^{\mathrm Q},0,0,0,0,0,M_{1,0}^{\mathrm Q},0\right]^{\mathrm T}.
\label{eq:new_quantized_joint_mean}
\end{equation}
Accordingly, the full covariance is expressed directly in terms of the focused-component statistics, leakage energy, and reciprocal-leakage moment in \eqref{eq:new_quantized_joint_covariance}.
\begin{figure*}[!b]
\centering
\rule{\textwidth}{0.4pt}
\vspace{-1.5ex}
\begin{equation}
\mathbf C_q^{\mathrm Q}
=
\begin{bmatrix}
\mathbf C_f^{\mathrm Q}&\mathbf 0_2&\mathbf 0_2&\mathbf C_{fd}^{\mathrm Q}\\
\mathbf 0_2&\frac{\eta_{\mathrm Q}}{2}\mathbf I_2&\frac{\tau_{\mathrm Q}}{2}\begin{bmatrix}1&0\\0&-1\end{bmatrix}&\mathbf 0_2\\
\mathbf 0_2&\frac{\tau_{\mathrm Q}}{2}\begin{bmatrix}1&0\\0&-1\end{bmatrix}&\frac{\eta_{\mathrm Q}}{2}\mathbf I_2&\mathbf 0_2\\
(\mathbf C_{fd}^{\mathrm Q})^{\mathrm T}&\mathbf 0_2&\mathbf 0_2&\mathbf C_f^{\mathrm Q}
\end{bmatrix}.
\label{eq:new_quantized_joint_covariance}
\end{equation}
\end{figure*}
The four block rows expose the physical composition of the quantized-phase model. Specifically, $\mathbf C_f^{\mathrm Q}$ describes the desired focusing uncertainty and the residual quadrature component, $\mathbf C_{fd}^{\mathrm Q}$ retains the cross-state dependence of the focused components, $\eta_{\mathrm Q}\mathbf I_2/2$ represents isotropic off-target leakage energy, and the blocks containing $\tau_{\mathrm Q}$ describe reciprocal leakage coupling. Compared with continuous phase control, phase quantization therefore affects both the useful focused component and the leakage geometry.
Hence,
\begin{equation}
\sum_{g=1}^G\mathbf q_g^{\mathrm Q}
\mathrel{\dot\sim}
\mathcal N\!\left(G\boldsymbol\mu_q^{\mathrm Q},G\mathbf C_q^{\mathrm Q}\right).
\label{eq:new_quantized_group_clt}
\end{equation}

\subsubsection{Continuous-Phase Benchmark}

For an index error $i\ne\widehat i$, the per-group real-valued contribution vector under continuous phase control is given by \eqref{eq:new_continuous_q}.
\begin{figure*}[!b]
\centering
\begin{equation}
\mathbf q^{\mathrm C}
=\left[
U_i^{i,\mathrm C},0,
\Re\{U_i^{\widehat i,\mathrm C}\},\Im\{U_i^{\widehat i,\mathrm C}\},
\Re\{U_{\widehat i}^{i,\mathrm C}\},\Im\{U_{\widehat i}^{i,\mathrm C}\},
U_{\widehat i}^{\widehat i,\mathrm C},0
\right]^{\mathrm T}.
\label{eq:new_continuous_q}
\end{equation}
\end{figure*}
The focused contributions $U_i^{i,\mathrm C}$ and $U_{\widehat i}^{\widehat i,\mathrm C}$ are real and nonnegative owing to continuous phase alignment.
Let $d\Lambda^{\mathrm C}$ denote the joint selection measure generated by the common source-link vector, the two target-link vectors, and the two continuous-phase arg-max events. For any function $\varphi$,
\begin{equation}
\begin{aligned}
\int\varphi\,d\Lambda^{\mathrm C}
&=\sum_{p=1}^P\sum_{q=1}^P
\int\varphi_{p,q}\,
\mathbf 1\{p=p_i^\star\}\mathbf 1\{q=p_{\widehat i}^\star\}\\
&\quad{}\times f_{\mathbf f,\mathbf h_i,\mathbf h_{\widehat i}}
d\mathbf f\,d\mathbf h_i\,d\mathbf h_{\widehat i}.
\end{aligned}
\label{eq:new_continuous_joint_selection_measure}
\end{equation}
The diagonal terms $p=q$ in \eqref{eq:new_continuous_joint_selection_measure} describe complete port overlap between the two receive-index states, whereas the terms $p\ne q$ describe distinct selected ports. This separation retains both overlap-induced coupling and the correlation between different fluid ports in the following moments.
Define the leakage energy on a key indexed antenna and the reciprocal-leakage pseudo-cross moment as
\begin{equation}
\eta_{\mathrm C}=\int|U_i^{\widehat i,\mathrm C}|^2d\Lambda^{\mathrm C},
\qquad
\tau_{\mathrm C}=\int U_i^{\widehat i,\mathrm C}U_{\widehat i}^{i,\mathrm C}d\Lambda^{\mathrm C}.
\label{eq:new_continuous_eta_tau}
\end{equation}
The quantity $\eta_{\mathrm C}$ is the average energy delivered to one indexed receive antenna when the FRIS is configured for the competing antenna. In contrast, $\tau_{\mathrm C}$ is a pseudo-cross moment rather than an energy; it retains the statistical coupling between the reciprocal off-target contributions $U_i^{\widehat i,\mathrm C}$ and $U_{\widehat i}^{i,\mathrm C}$. The real reference-port correlations and continuous phase compensation imply $\tau_{\mathrm C}\in\mathbb R$.
The mean is
\begin{equation}
\boldsymbol\mu_q^{\mathrm C}
=\left[M_1^{\mathrm C},0,0,0,0,0,M_1^{\mathrm C},0\right]^{\mathrm T}.
\label{eq:new_continuous_joint_mean}
\end{equation}
The covariance has the structured form in \eqref{eq:new_continuous_joint_covariance}, in which every nonzero block has a direct physical interpretation.
\begin{figure*}[!b]
\centering
\begin{equation}
\mathbf C_q^{\mathrm C}
=
\begin{bmatrix}
\sigma_{\mathrm C}^2&0&0&0&0&0&\xi_{\mathrm C}&0\\
0&0&0&0&0&0&0&0\\
0&0&\eta_{\mathrm C}/2&0&\tau_{\mathrm C}/2&0&0&0\\
0&0&0&\eta_{\mathrm C}/2&0&-\tau_{\mathrm C}/2&0&0\\
0&0&\tau_{\mathrm C}/2&0&\eta_{\mathrm C}/2&0&0&0\\
0&0&0&-\tau_{\mathrm C}/2&0&\eta_{\mathrm C}/2&0&0\\
\xi_{\mathrm C}&0&0&0&0&0&\sigma_{\mathrm C}^2&0\\
0&0&0&0&0&0&0&0
\end{bmatrix}.
\label{eq:new_continuous_joint_covariance}
\end{equation}
\end{figure*}
In \eqref{eq:new_continuous_joint_covariance}, the zero imaginary variances of $U_i^{i,\mathrm C}$ and $U_{\widehat i}^{\widehat i,\mathrm C}$ follow from perfect phase compensation. The terms $\sigma_{\mathrm C}^2$ describe coherent focused-gain fluctuations, $\eta_{\mathrm C}/2$ gives the real- or imaginary-component energy leaked to the other indexed antenna, $\xi_{\mathrm C}$ couples the two state-dependent focused gains, and $\tau_{\mathrm C}$ couples the reciprocal leakage paths. The matrix therefore separates the desired focusing statistics from the leakage and overlap-induced cross-state statistics.
Since the $G$ groups are independent, the multivariate central limit theorem (CLT) yields the following approximation in distribution:
\begin{equation}
\sum_{g=1}^G\mathbf q_g^{\mathrm C}
\mathrel{\dot\sim}
\mathcal N\!\left(G\boldsymbol\mu_q^{\mathrm C},G\mathbf C_q^{\mathrm C}\right).
\label{eq:new_continuous_group_clt}
\end{equation}

\subsection{Distance Transforms}

\subsubsection{Receive-Index Errors}

Let $b\in\{\mathrm C,\mathrm Q\}$. For $i\ne\widehat i$, the real error vector on the two key receive antennas is
\begin{equation}
\mathbf v^b
=\mathbf T_{m,\widehat m}\sum_{g=1}^G\mathbf q_g^b,
\label{eq:new_index_error_vector}
\end{equation}
where $\mathbf T_{m,\widehat m}$ is given by \eqref{eq:new_index_transform_matrix}.
\begin{figure*}[!b]
\centering
\begin{equation}
\mathbf T_{m,\widehat m}
=
\begin{bmatrix}
s_{m,\mathrm R}&-s_{m,\mathrm I}&-s_{\widehat m,\mathrm R}&s_{\widehat m,\mathrm I}&0&0&0&0\\
s_{m,\mathrm I}&s_{m,\mathrm R}&-s_{\widehat m,\mathrm I}&-s_{\widehat m,\mathrm R}&0&0&0&0\\
0&0&0&0&s_{m,\mathrm R}&-s_{m,\mathrm I}&-s_{\widehat m,\mathrm R}&s_{\widehat m,\mathrm I}\\
0&0&0&0&s_{m,\mathrm I}&s_{m,\mathrm R}&-s_{\widehat m,\mathrm I}&-s_{\widehat m,\mathrm R}
\end{bmatrix}.
\label{eq:new_index_transform_matrix}
\end{equation}
\end{figure*}
Thus,
\begin{equation}
\boldsymbol\mu_v^b
=G\mathbf T_{m,\widehat m}\boldsymbol\mu_q^b,
\qquad
\mathbf C_v^b
=G\mathbf T_{m,\widehat m}\mathbf C_q^b\mathbf T_{m,\widehat m}^{\mathrm T}.
\label{eq:new_index_transformed_moments}
\end{equation}
The Laplace transform of the key-branch distance is
\begin{equation}
\begin{aligned}
\mathcal L_{\|\mathbf v^b\|^2}(t)
&=\det(\mathbf I_4+2t\mathbf C_v^b)^{-1/2}\\
&\quad{}\times
\exp\!\left[-t(\boldsymbol\mu_v^b)^{\mathrm T}
(\mathbf I_4+2t\mathbf C_v^b)^{-1}\boldsymbol\mu_v^b\right].
\end{aligned}
\label{eq:new_index_key_laplace}
\end{equation}
The term $\|\mathbf v^b\|^2$ is the distance contributed by the two antennas whose receive-index labels are confused. It jointly captures the desired focused components, reciprocal off-target leakage, and their cross moments on the two indexed antennas.
For a remaining receive antenna $\ell\notin\{i,\widehat i\}$, define the cross-state leakage coherence as
\begin{equation}
\chi_b=\mathbb E\!\left[U_{\ell}^{i,b}(U_{\ell}^{\widehat i,b})^*\right],
\label{eq:new_remaining_cross_moment}
\end{equation}
To distinguish their physical roles, let $\eta_{b,\mathrm k}=\mathbb E[|U_i^{\widehat i,b}|^2]$ denote the leakage energy on the competing key antenna and let $\eta_{b,\mathrm o}=\mathbb E[|U_{\ell}^{i,b}|^2]$ denote that on an ordinary non-target antenna. The receive channel on a branch not targeted by a given state does not affect that state's port-selection event. Since $\mathbb E[|h_{r,p}|^2]=1$ for every receive antenna and the selection indicators form a partition, these two energies coincide:
\begin{equation}
\eta_{b,\mathrm k}=\eta_{b,\mathrm o}\equiv\eta_b
=\sum_{p=1}^P\mathbb E\!\left[|f_p|^2\mathbf 1\{p_i^{\star,b}=p\}\right].
\label{eq:new_remaining_leakage_energy_explicit}
\end{equation}
Thus, the same scalar $\eta_b$ enters the key-branch covariance matrix and the ordinary-branch distance factor, while the two labels expose their different physical locations. To evaluate \eqref{eq:new_remaining_cross_moment}, condition on the source-link vector. Proper circular symmetry of each target receive-channel vector leaves its port-selection metric unchanged under a common phase rotation. The corresponding continuous phase shifts by the opposite rotation, whereas, for finite-bit control, a rotation by one quantization interval $\Delta_Q$ cyclically shifts the selected phase without changing the wrapped quantization error or the selected port. Consequently, for $b\in\{\mathrm C,\mathrm Q\}$,
\begin{equation}
\mathbb E\!\left[e^{\mathrm j\theta_i^{\star,b}}\mathbf 1\{p_i^{\star,b}=p\}\mid\mathbf f\right]=0,
\qquad p=1,\ldots,P,
\label{eq:new_phase_selection_zero_mean}
\end{equation}
where the quantized-phase result holds for $Q\geq1$. Since the receive-channel vectors associated with different antennas are independent, \eqref{eq:new_phase_selection_zero_mean} gives
\begin{equation}
\chi_{\mathrm C}=\chi_{\mathrm Q}=0.
\label{eq:new_remaining_cross_moment_zero}
\end{equation}
Hence, the average distance contributed by one ordinary non-target branch is $(|s_m|^2+|s_{\widehat m}|^2)\eta_b$. Combining this contribution with the joint distance on the two indexed antennas gives
\begin{equation}
\mathcal L_{\mathcal D}^b(t)
\simeq
\mathcal L_{\|\mathbf v^b\|^2}(t)
\Big(1+tG(|s_m|^2+|s_{\widehat m}|^2)\eta_b\Big)^{-(N_{\mathrm r}-2)}.
\label{eq:new_index_distance_transform}
\end{equation}
Equation~\eqref{eq:new_index_distance_transform} applies to $i\ne\widehat i$.

\subsubsection{Symbol-Only Errors}

When $i=\widehat i$, define the symbol-difference transform directly as
\begin{equation}
\mathbf T_{m,\widehat m}^{\Delta}
=\begin{bmatrix}
\Re\{s_m-s_{\widehat m}\}&-\Im\{s_m-s_{\widehat m}\}\\
\Im\{s_m-s_{\widehat m}\}&\Re\{s_m-s_{\widehat m}\}
\end{bmatrix}.
\label{eq:new_symbol_transform_matrix}
\end{equation}
For continuous phase, define
\begin{equation}
\boldsymbol\mu_f^{\mathrm C}
=\begin{bmatrix}M_1^{\mathrm C}\\0\end{bmatrix},
\qquad
\mathbf C_f^{\mathrm C}
=\begin{bmatrix}M_2^{\mathrm C}-(M_1^{\mathrm C})^2&0\\0&0\end{bmatrix},
\label{eq:new_continuous_focused_2d_moments}
\end{equation}
while $\boldsymbol\mu_f^{\mathrm Q}$ and $\mathbf C_f^{\mathrm Q}$ are given by \eqref{eq:new_quantized_focused_mean_cov_symmetric}. For either $b\in\{\mathrm C,\mathrm Q\}$, define
\begin{equation}
\boldsymbol\mu_d^b
=G\mathbf T_{m,\widehat m}^{\Delta}\boldsymbol\mu_f^b,
\qquad
\mathbf C_d^b
=G\mathbf T_{m,\widehat m}^{\Delta}\mathbf C_f^b(\mathbf T_{m,\widehat m}^{\Delta})^{\mathrm T}.
\label{eq:new_symbol_transformed_moments}
\end{equation}
Here, the first two factors below describe the symbol distance carried by the coherently focused branch, including the quadrature uncertainty under quantized control. The final factor collects the same symbol difference observed through the $N_{\mathrm r}-1$ non-target leakage branches. Although such leakage reduces the energy contrast used by a greedy index detector, full-vector ML detection can exploit these additional observations as part of the Euclidean distance.
The symbol-error distance transform is
\begin{equation}
\begin{aligned}
\mathcal L_{\mathcal D}^b(t)
&\simeq
\det(\mathbf I_2+2t\mathbf C_d^b)^{-1/2}\\
&\quad{}\times
\exp\!\left[-t(\boldsymbol\mu_d^b)^{\mathrm T}
(\mathbf I_2+2t\mathbf C_d^b)^{-1}\boldsymbol\mu_d^b\right]\\
&\quad{}\times
\left(1+tG|s_m-s_{\widehat m}|^2\eta_b\right)^{-(N_{\mathrm r}-1)}.
\end{aligned}
\label{eq:new_symbol_distance_transform}
\end{equation}
Equation~\eqref{eq:new_symbol_distance_transform} applies to $i=\widehat i$ and $m\ne\widehat m$. FPS-RIM-IO has $M=1$ and therefore contains no symbol-only error events.

\subsection{BER Union Bound}

Substituting \eqref{eq:new_index_distance_transform} or \eqref{eq:new_symbol_distance_transform} into \eqref{eq:new_craig_pep} gives the UPEP for continuous or quantized phase control. Let $\mathbf b_{i,m}$ be the bit label of transmit state $(i,s_m)$, let $d_{\mathrm H}(\cdot,\cdot)$ denote the Hamming distance, and let $B=\log_2(N_{\mathrm r}M)$. The BER union bound is
\begin{equation}
P_{\mathrm b}^b
\le
\frac{1}{N_{\mathrm r}MB}
\sum_{i=1}^{N_{\mathrm r}}\sum_{m=1}^M
\sum_{\widehat i=1}^{N_{\mathrm r}}\sum_{\widehat m=1}^M
d_{\mathrm H}(\mathbf b_{i,m},\mathbf b_{\widehat i,\widehat m})
\overline P_{i,\widehat i}^{m,\widehat m},
\label{eq:new_ber_union_bound}
\end{equation}
where the identical pair $(i,m)=(\widehat i,\widehat m)$ is excluded and $b=\mathrm C$ or $b=\mathrm Q$ denotes continuous or quantized phase control, respectively.

\section{Numerical Results and Discussion}
\label{sec:numerical_results}

This section evaluates the BER of FPS-RIM-IO and FPS-RIM-SA and compares the simulations with the analysis in Section~\ref{section_performance_analysis}. Unless otherwise specified, the Tx--FRIS and FRIS--Rx links follow the reference-port correlated Rayleigh model in Section~\ref{section_system_model}. Candidate ports in the same group follow the Jakes correlation model, while ports in different groups are uncorrelated because their separations exceed one half-wavelength. All average channel powers are normalized to unity, and the SNR is $E_s/N_0$. FPS-RIM-SA uses Gray-labeled PSK symbols with unit average energy. Channel acquisition is assumed to have been completed before data transmission; the available CSI is used for FRIS configuration and coherent detection. The results consider the number of candidate ports per group, phase resolution, number of groups, group aperture, and detector list size. The analytical curves are obtained from Section~\ref{section_performance_analysis}.

\begin{figure}[!t]
  \centering
  \includegraphics[width=0.85\columnwidth]{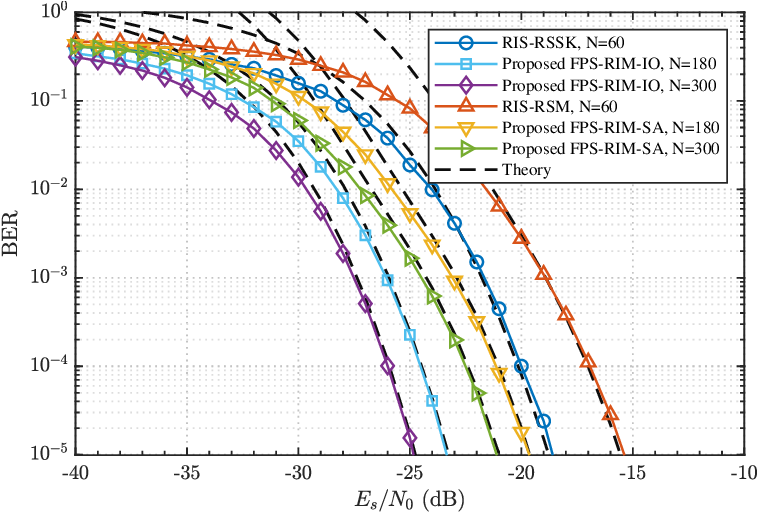}
  \caption{BER performance of the proposed FPS-RIM schemes and their conventional RIS counterparts for different numbers of candidate fluid ports per group under $2$-bit phase control.}
  \label{fig:result_1}
\end{figure}

Fig.~\ref{fig:result_1} compares FPS-RIM-IO and FPS-RIM-SA for different numbers of candidate ports per group under $2$-bit phase control. The parameters are $N_{\mathrm r}=8$, $G=60$, $W_{\mathrm g}=0.5\lambda$, and $P\in\{1,3,5\}$, giving $N=GP\in\{60,180,300\}$ candidate ports. One port is active in each group, so the number of active ports remains $60$. The IO mode conveys only the receive-index bits, whereas the SA mode also employs $8$-PSK. For $P=1$, no port selection is available and the two modes reduce to RIS-RSSK and RIS-RSM~\cite{Basar2020RIS_IM}, respectively. For $P=3$ and $P=5$, the adjacent-port spacings are $0.25\lambda$ and $0.125\lambda$. At a BER of $10^{-4}$, these two configurations provide gains of $4.5$ and $6.0$~dB over RIS-RSSK for FPS-RIM-IO, and $4.2$ and $5.6$~dB over RIS-RSM for FPS-RIM-SA. Because the number of active ports is fixed, the gains result from selecting a stronger cascaded channel within each group rather than combining more active paths. The similar gains of the two modes indicate that symbol transmission does not remove the port-selection benefit. The analytical and simulated BER curves remain close throughout the plotted range.

\begin{figure}[!t]
  \centering
  \includegraphics[width=0.85\columnwidth]{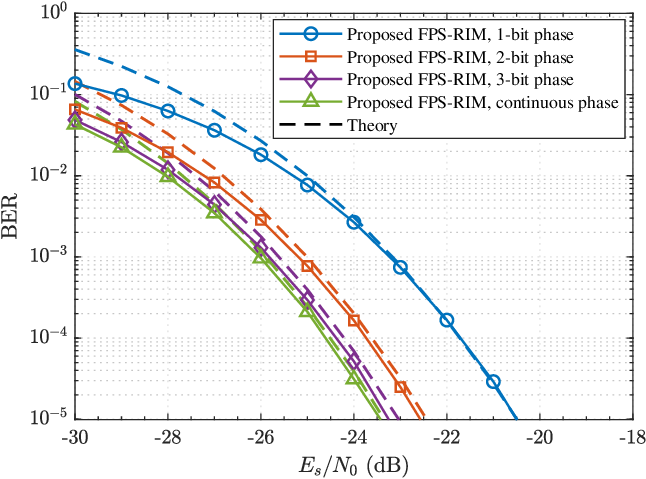}
  \caption{BER performance of the proposed FPS-RIM-SA scheme with continuous-
  and finite-resolution phase control.}
  \label{fig:result_2}
\end{figure}

Fig.~\ref{fig:result_2} compares FPS-RIM-SA under continuous and finite-resolution phase control for $N_{\mathrm r}=4$, $M=4$, $G=50$, $P=5$, and $W_{\mathrm g}=0.5\lambda$, corresponding to $N=250$ candidate ports. Relative to continuous phase control, the $1$-, $2$-, and $3$-bit configurations require an additional $2.9$, $0.9$, and $0.2$~dB, respectively, at a BER of $10^{-4}$. A finer phase resolution reduces both the loss of the in-phase focusing component and the quadrature leakage. The $3$-bit result is consequently close to the continuous-phase benchmark. The analytical curves reproduce this trend and closely follow the simulations for all four phase resolutions.

\begin{figure}[!t]
  \centering
  \includegraphics[width=0.85\columnwidth]{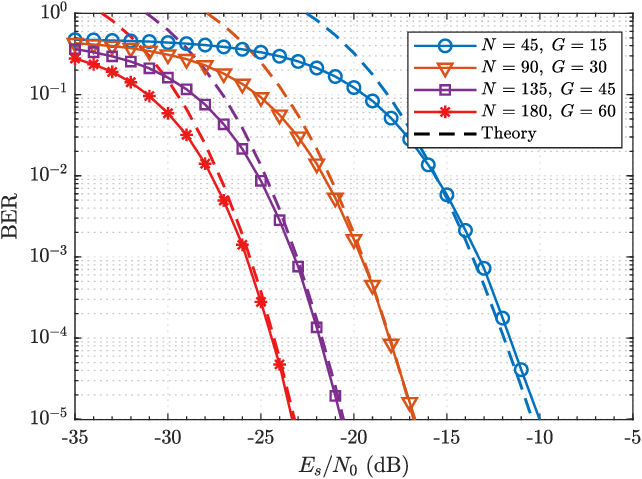}
  \caption{BER performance of the proposed FPS-RIM-SA scheme for different numbers of fluid-port groups under $3$-bit phase control.}
  \label{fig:result_3}
\end{figure}

Fig.~\ref{fig:result_3} compares FPS-RIM-SA for $G\in\{15,30,45,60\}$ with $N_{\mathrm r}=8$, $M=4$, $P=3$, $W_{\mathrm g}=0.5\lambda$, and $Q=3$. The corresponding numbers of candidate ports are $N\in\{45,90,135,180\}$. Because one port is active in each group, a larger $G$ adds independently selected and phase-aligned contributions at the intended antenna. The nonfocused components are not coherently aligned, and the BER decreases as $G$ increases. The gap between analysis and simulation also narrows with $G$, since the equivalent channel is then formed from more independent group contributions and the multivariate CLT approximation becomes more accurate.

\begin{figure}[!t]
  \centering
  \includegraphics[width=0.85\columnwidth]{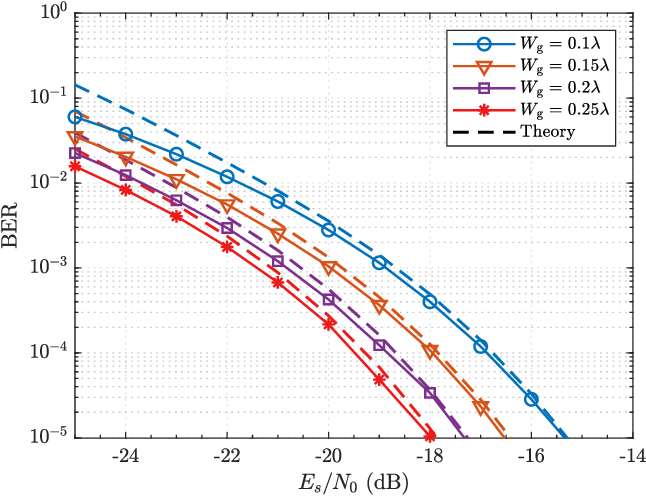}
  \caption{BER performance of the proposed FPS-RIM-SA scheme for different fluid-port group apertures under continuous phase control.}
  \label{fig:result_4}
\end{figure}

Fig.~\ref{fig:result_4} examines the group aperture for FPS-RIM-SA with $N_{\mathrm r}=4$, $M=8$, $G=40$, $P=8$, and continuous phase control. The aperture $W_{\mathrm g}$ ranges from $0.1\lambda$ to $0.25\lambda$. Continuous phase control removes phase-quantization loss from this comparison, while the total number of candidate ports remains $N=320$. At a BER of $10^{-4}$, increasing $W_{\mathrm g}$ from $0.1\lambda$ to $0.15\lambda$, $0.2\lambda$, and $0.25\lambda$ yields gains of $1.1$, $2.0$, and $2.6$~dB, respectively. The wider aperture increases the spacing between adjacent ports and reduces their spatial correlation. It therefore provides greater channel variation for port selection and lowers the probability of simultaneous deep fading within a group. The analytical curves follow the simulations across the four apertures.

\begin{figure}[!t]
  \centering
  \includegraphics[width=0.85\columnwidth]{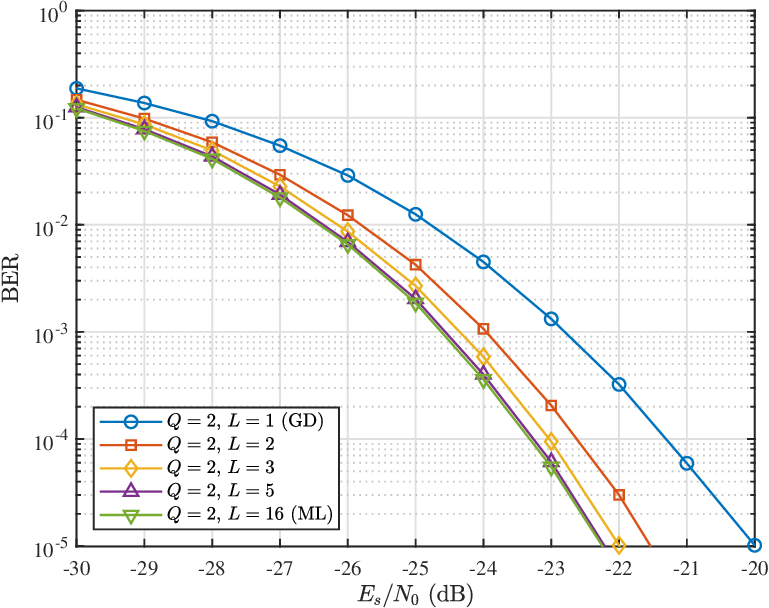}
  \caption{BER performance of the proposed FPS-RIM-SA scheme with the low-complexity Top-$L$ detector under $2$-bit phase control.}
  \label{fig:result_5}
\end{figure}

Fig.~\ref{fig:result_5} compares the low-complexity Top-$L$ detector for $N_{\mathrm r}=16$, $M=4$, $G=50$, $P=4$, $W_{\mathrm g}=0.5\lambda$, and $Q=2$. The detector ranks the receive-index states by received energy and evaluates the joint Euclidean distance over the retained $LM$ index-symbol hypotheses. The cases $L=1$ and $L=16$ correspond to GD and ML detection, respectively. At a BER of $10^{-4}$, increasing $L$ from $1$ to $2$, $3$, and $5$ yields gains of $1.3$, $1.7$, and $2.0$~dB. The loss relative to ML is $0.7$~dB for $L=2$, $0.3$~dB for $L=3$, and $0.1$~dB for $L=5$. FRIS focusing generally places the transmitted receive-index state among the strongest energy candidates, while a list with $L>1$ prevents a single incorrect energy ranking from fixing the final decision. With $L=5$, the low-complexity detector evaluates $20$ rather than $64$ joint hypotheses and remains within $0.1$~dB of ML detection.

\section{Conclusion}
\label{sec:conclusion}

This paper investigated FPS-RIM under finite-resolution FRIS phase control over spatially correlated cascaded channels. The IO and SA modes convey information through receive-antenna indices alone or jointly with constellation symbols. A low-complexity Top-$L$ detector restricts joint detection to the receive-index states retained by energy ranking. For the BER analysis, we derived selection-aware joint moments that account for phase quantization, fluid-port selection, and state-dependent port overlap. These moments yield correlated Gaussian distance models, UPEP, and union-bound BER expressions, with continuous phase control serving as a benchmark. Numerical results showed that larger candidate-port sets, higher phase resolution, and lower intra-group correlation improve BER. The Top-$L$ detector approached ML performance with fewer hypotheses, while the analytical curves closely followed the simulations.

\bibliographystyle{IEEEtran}
\bibliography{Reference}

\end{document}